\documentstyle[12pt,epsfig]{article}

\def\bb{b \bar{b}} 
\def\cc{c \bar{c}}

\begin{document}  
\vspace*{-2cm}  
\renewcommand{\thefootnote}{\fnsymbol{footnote}}  
\begin{flushright}  
hep-ph/9907473\\
DTP/99/82\\  
July 1999\\  
\end{flushright}  
\vskip 65pt  
\begin{center}  
{\Large \bf Precise measurement of $\Gamma ( H \longrightarrow \gamma \gamma)$
at a PLC and theoretical consequences} \\
\vspace{1.2cm} 
{\bf  
Michael Melles\footnote{Michael.Melles@durham.ac.uk}  
\footnote{Research supported by the EU Fourth Framework Programme
	`Training and Mobility of Researchers' through a Marie Curie
	Fellowship.}
}\\  
\vspace{10pt}  
Department of Physics, University of Durham,  
Durham DH1 3LE, U.K.\\  

\vspace{70pt}  
\begin{abstract}
With the LEP II Higgs search approaching exclusion limits
on low values of $\tan \beta \sim 2$ it
becomes increasingly important to investigate physical quantities sensitive
to large masses of a pseudoscalar Higgs mass. This regime is difficult  
and over a large range of $\tan \beta$ impossible to cover at the LHC
proton proton collider. In this paper we focus on the achievable
statistical precision of the Higgs decay into two photons at a future $\gamma
\gamma$ collider (PLC) in the MSSM mass range below 130 GeV. 
The MSSM and SM predictions for $\Gamma ( H \longrightarrow \gamma \gamma)$
can differ by up to 10 \% even in the decoupling limit of large $m_A$. 
We summarize recent progress in both the theoretical understanding
of the background process $\gamma \gamma \longrightarrow q \overline{q}$,
$q=\{b,c\}$, and in the expected detector performance allow for a high 
accuracy of the lightest MSSM or SM Higgs boson decay into a $b \overline{b}$
pair. 
We find that for optimized but still realistic detector and accelerator
assumptions, statistically a 1.4\% accuracy is feasible after about
four years of collecting data for a Higgs boson mass which excludes $\tan \beta
<2$.
\end{abstract}
\end{center}  
\vskip12pt

\setcounter{footnote}{0}  
\renewcommand{\thefootnote}{\arabic{footnote}}  
  
\vfill  
\clearpage  
\setcounter{page}{1}  
\pagestyle{plain} 
 
\section{Introduction} 

Uncovering the origin of particle masses is set to dominate the coming decade in 
high energy physics. It is commonly assumed that the Higgs mechanism \cite{h}
gives rise to all vector boson and fermion masses present in the Standard
Model (SM) and indirect experimental evidence points to the existence
of a light Higgs, possibly in the range predicted by the minimal
supersymmetric SM (MSSM) \cite{ghkd}. The MSSM is attractive to theoreticians for
several reasons, mostly, however, because supersymmetry stabilizes the 
quadratic 
divergences of scalars when the theory is extrapolated to GUT-scale energies.
If one believes in such vast extrapolations, the SM-Higgs must be in a window
between roughly 130-180 GeV due to restrictions on vacuum
stability and the perturbative framework respectively \cite{sher,hk}, not considering fine tuning 
(hierarchy) problems. Alternatively, a lighter Higgs mass would indicate the scale at which to
expect new physics \cite{hk}.

Supersymmetry and the requirement of
anomaly cancellation both require al least two complex Higgs doublets in the
MSSM leading
to five physical degrees of freedom, two neutral CP-even (h,H), one neutral 
CP-odd (A) and two charged Higgs bosons (H$^\pm$) \cite{ghkd}. In contrast to a general
two doublet Higgs model (2DHM), the MSSM Higgs sector has (at tree level) only
two free parameters, commonly chosen as the ratio of the vacuum expectation 
values of up-and down-type Higgs bosons, $\tan \beta$, and the mass of the
pseudoscalar Higgs, $m_A$.
The lightest neutral scalar, h, in this model
must be below 130 GeV \cite{hhw}. 

Any supersymmetric extension of the SM must, in order to be phenomenologically 
viable, contain terms in the Lagrangian which break it. Commonly one introduces
so called soft SUSY-breaking terms which can be thought to originate from
supergravity or gauge mediation for instance \cite{sm}. In typical `sugra'
scenarios, the squark and gaugino scales are each degenerate at the SUSY-GUT scale
and mass terms then evolve down to electroweak (EW) energies. EW-symmetry
breaking in these models is then brought about by the Higgs boson mass parameter $m_2$
developing a negative vacuum expectation value
\cite{bbo}.

While {\it a priori} a multitude of possibilities exists considering all the
undetermined parameters of models beyond the SM, plausible SUSY scenarios often
predict heavy pseudoscalar masses, $m_A>400$ GeV \cite{bbo}. At the LHC proton
proton collider the sensitivity to these values of $m_A$ is very limited.
A negative Higgs search at LEP II would lead to a lower bound 
$\tan \beta > 2$ in the MSSM. 
Values of $\tan \beta >3$ imply that for
pseudoscalar masses above 250 GeV the LHC would not be able to provide
information on $m_A$ for intermediate values of $\tan \beta$ and above
500 GeV, only for very large $\tan \beta$ the $\tau^+ \tau^-$ decay could
be utilized \cite{s}. 

In this context it is therefore of considerable interest to study observables
which possess a large enough sensitivity to $m_A$ in the decoupling limit
compared to the SM prediction. The partial Higgs width $\Gamma ( H \longrightarrow
\gamma \gamma)$, measured at the $\gamma \gamma$ Compton-backscattered option
of a future linear $e^\pm$ collider, is such a quantity \cite{bbc}.
The decoupling limit of the diphoton Higgs width was studied in Ref. \cite{ddhi}
and it was found that the MSSM and SM predictions can differ by up to
10 \% in this limit. 
It was recently demonstrated in Ref. \cite{msk,m} that using conservative assumptions
an accuracy of 2\% is feasible at a PLC. In the next section we briefly review the
status of radiative corrections to both the signal (S) $\gamma \gamma \longrightarrow
H \longrightarrow b \overline{b}$ and background (BG) process
$\gamma \gamma \longrightarrow q \overline{q}$, $q=\{b,c\}$ and then present new
Monte Carlo results assuming a slightly more optimistic (but still
realistic) detector performance
as in \cite{msk}. The results are summarized in section \ref{sec:con} in the
above context of achievable precision measurements at a future photon photon
collider.

\section{Radiative Corrections}

\begin{center}
\begin{figure}
\centering
\epsfig{file=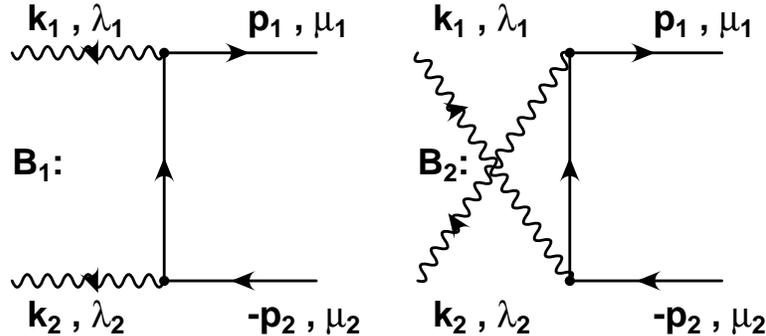,width=10cm}
\caption{The Born process in polarized $\gamma + \gamma (J_z=0) \longrightarrow
q + \overline{q}$. QCD radiative corrections lead to large non-Sudakov double logarithms which
need to be resummed.}
\label{fig:born}
\end{figure}
\end{center}
\begin{center}
\begin{figure}
\centering
\epsfig{file=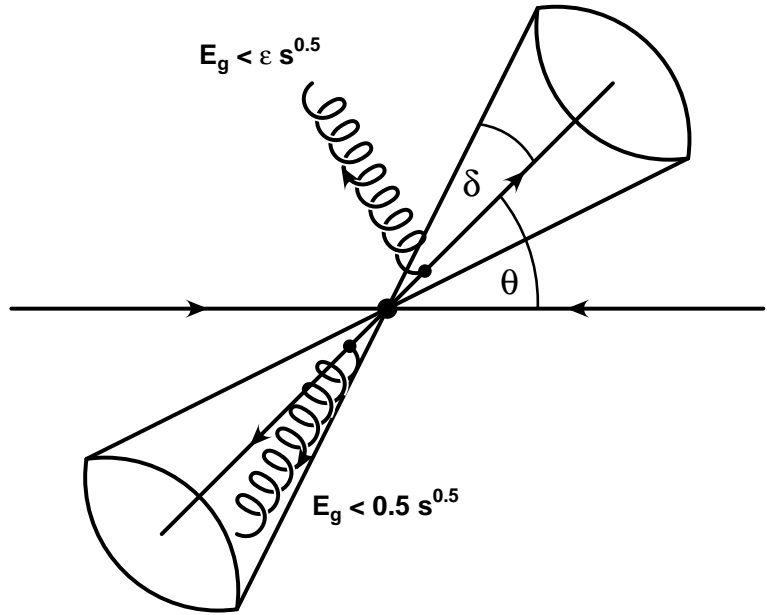,width=10cm}
\caption{The parameters of the Sterman-Weinberg two-jet definition used in this
work. Inside an angular cone of size $\delta$ arbitrary hard gluon bremsstrahlung
is included. Radiation outside this cone is only permitted if the gluon energy
is below a certain fraction ($\epsilon$) of the incident center of mass energy.
The thrust angle is denoted by $\theta$.}
\label{fig:jet}
\end{figure}
\end{center}

There has been considerable progress in the theoretical understanding of the
BG to the intermediate mass Higgs boson decay into $b \overline{b}$ recently.
The Born cross section in Fig. \ref{fig:born} for the $J_z=0$ channel is suppressed by $\frac{m_q^2}{
s}$ relative to the $J_z=\pm2$ which means that by ensuring a high degree of
polarization of the incident photons\footnote{Ratios of $J_0/J_2 =20..50$ are
feasible in presently considered designs, e.g. Ref. \cite{t}.} one can 
{\it simultaneously} enhance the signal and suppress the background.
QCD radiative corrections can remove this suppression, however, and large 
bremsstrahlung
and double logarithmic corrections need to be taken into account. 
\begin{center}
\begin{figure}
\centering
\epsfig{file=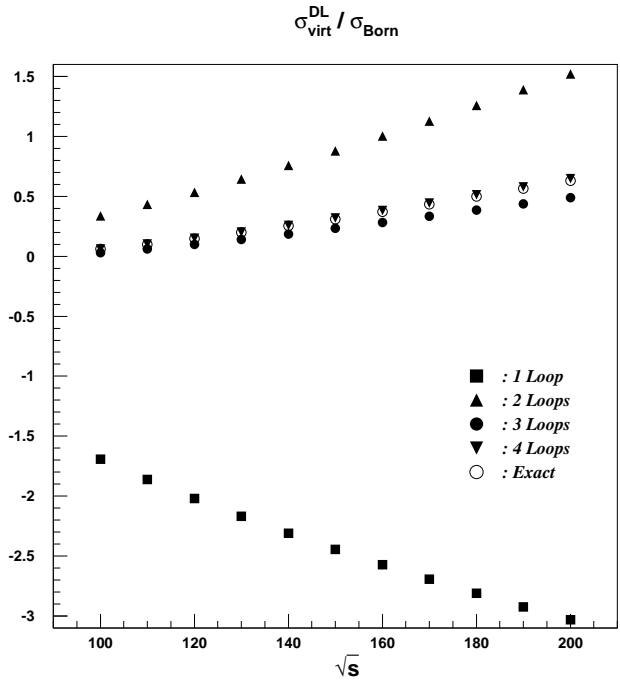,height=10cm}
\caption{The size of the virtual non-Sudakov double logarithmic (DL) contribution relative to
the Born cross section through four loops. The exact DL result (open circles) is given
by the all orders resummation according to Ref. \cite{ms1} and is in very good agreement
with the four loop approximation given in Ref. \cite{ms2}. The huge one and
loop contributions can be seen to lead to physically distorted results.}
\label{fig:4l}
\end{figure}
\end{center}
In Ref. \cite{jt} the exact one loop corrections to $\gamma \gamma 
\longrightarrow q \overline{q}$ were calculated and the largest virtual correction
was contained in novel non-Sudakov double logarithms. For some choices of
the invariant mass cutoff $y_{cut}$ even a negative cross section was obtained
in this approximation. The authors of Ref. \cite{fkm} elucidated the physical
nature of the novel double logarithms and performed a two loop calculation
in the DL-approximation. The results restored positivity to the physical
cross section. In Ref. \cite{ms1}, three loop DL-results were presented which
revealed a factorization of Sudakov and non-Sudakov DL's and led to the
all orders resummation of all DL in form of a confluent hypergeometric function
$_2F_2$. The general form of the expression is $\sigma_{DL}=\sigma_{Born}(1+
{\cal F}_{DL}) \exp({\cal F}_{Sud})$.
Fig \ref{fig:4l} demonstrates that at least four loops on the
cross section level are required to achieve a converged DL result \cite{ms2}. 
\begin{center}
\begin{figure}
\centering
\epsfig{file=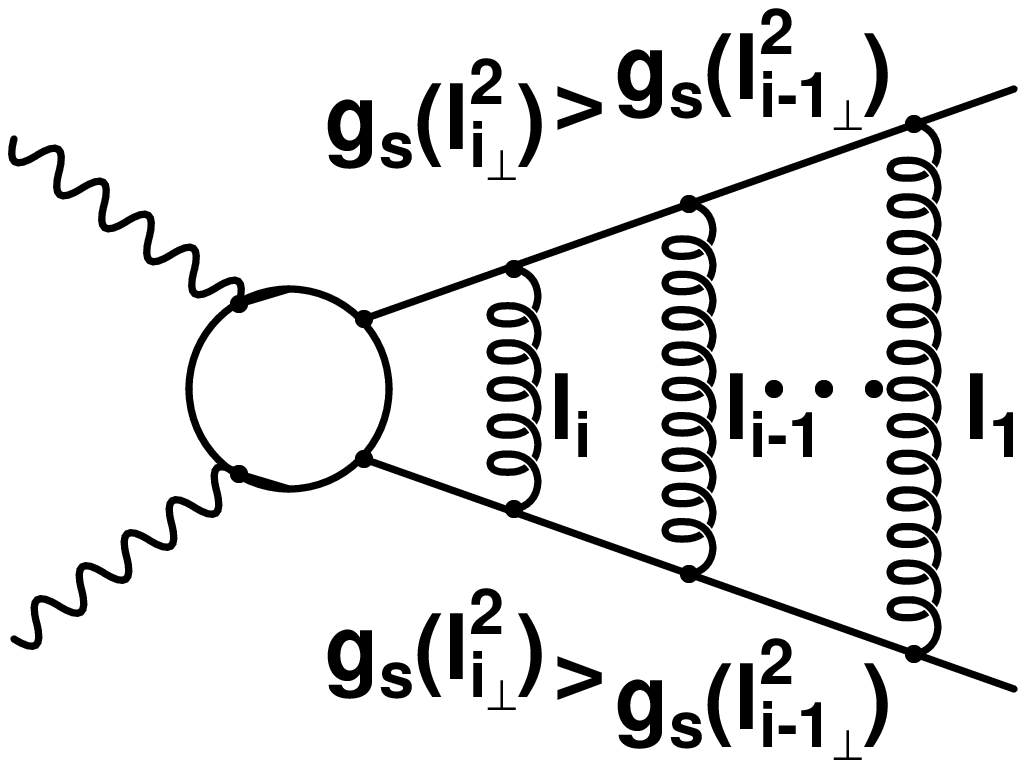,width=10cm}
\vspace{1cm}\\
\epsfig{file=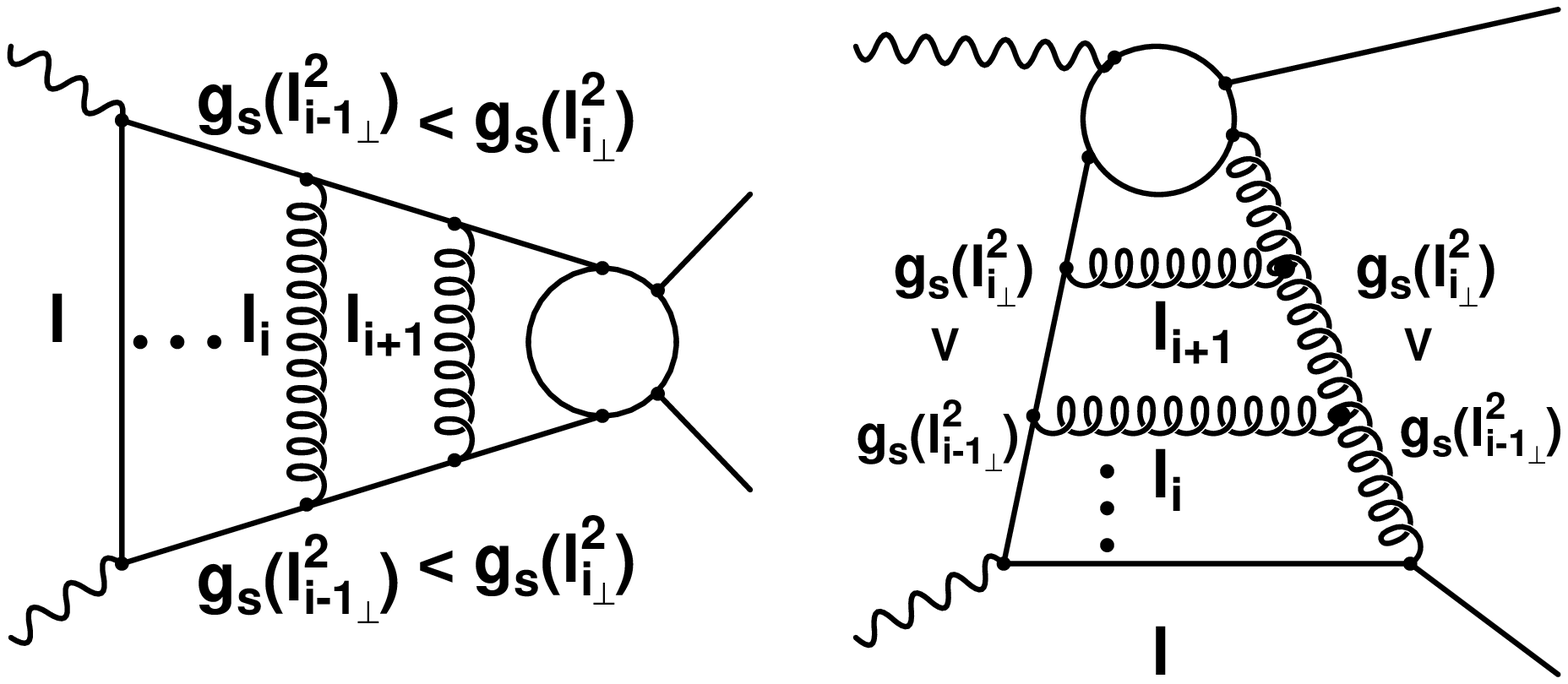,width=15cm}
\caption{The upper plot shows a schematic Feynman diagram leading to the Sudakov
double logarithms in the process $\gamma \gamma (J_z=0) \longrightarrow q
\overline{q}$ with $i$ gluon insertions.
The blob denotes a hard momentum going through the omitted propagator
in the DL-phase space.
Crossed diagrams lead to a different
ordering of the Sudakov variables with all resulting $C_A$ terms canceling the
DL-contributions from three gluon insertions \cite{ms1}.
The scale of the coupling $\alpha_s=\frac{g_s^2}{4
\pi}$ is indicated at the vertices and explicitly taken into
account in this work.
The lower row depicts schematic Feynman diagrams leading to the 
renormalization group improved hard (non-Sudakov)
double logarithms in that process.
The topology on
the left-hand diagram is Abelian like, and the one on the right is
non-Abelian beyond one loop.}
\label{fig:rg}
\end{figure}
\end{center}
At this
point the scale of the QCD-coupling is still unrestrained and differs by
more than a factor of two in-between the physical scales of the problem,
$m_q$ and $m_H$. This uncertainty was removed in Ref. \cite{ms3} by introducing
a running coupling $\alpha_s ({\bf l^2_\perp})$ into each loop integration (see Fig. \ref{fig:rg}), 
where $l_\perp$ denotes the perpendicular Sudakov loop momentum. 
\begin{center}
\begin{figure}
\centering
\epsfig{file=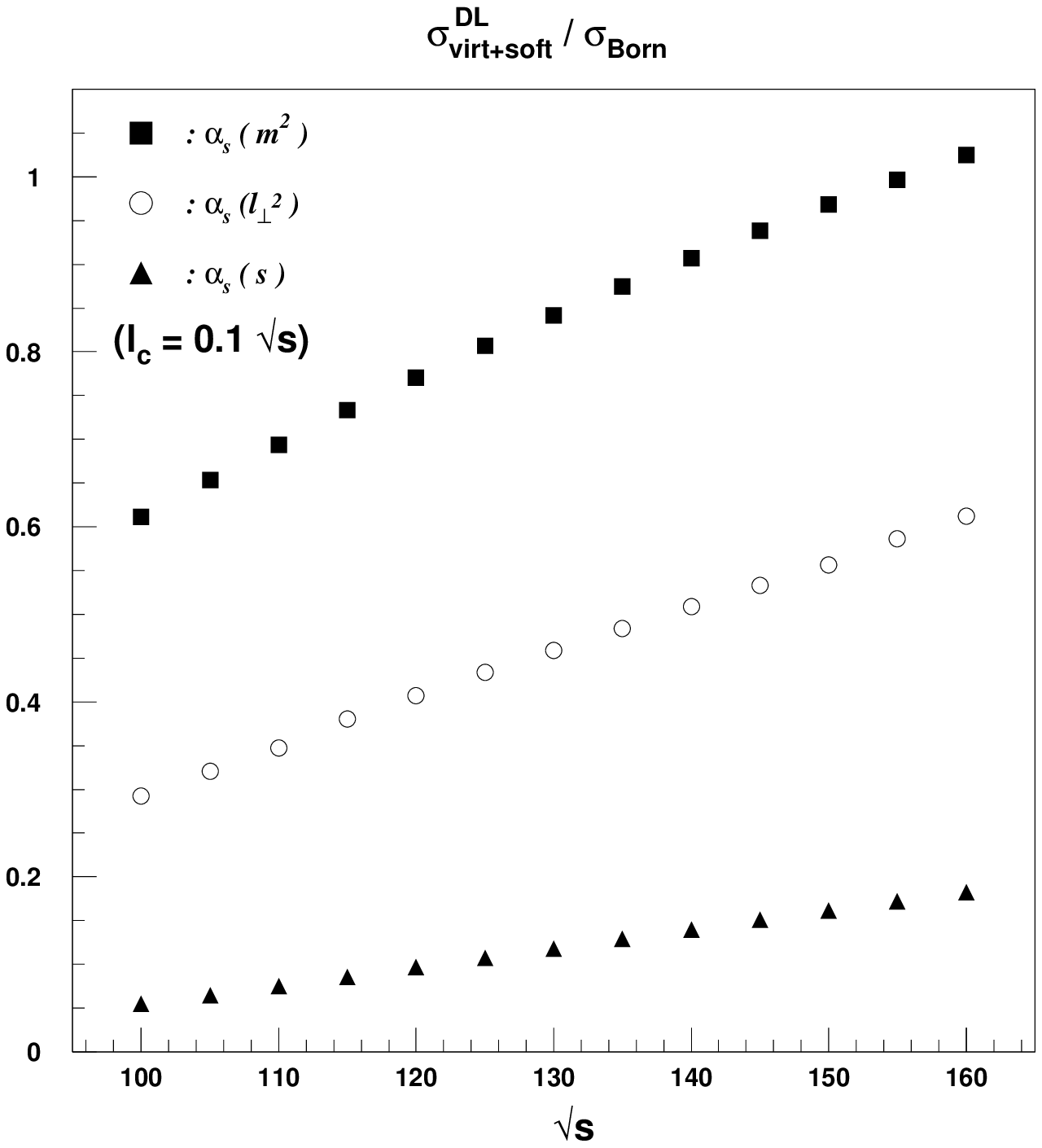,width=10cm}
\vspace{-1cm} \\
\epsfig{file=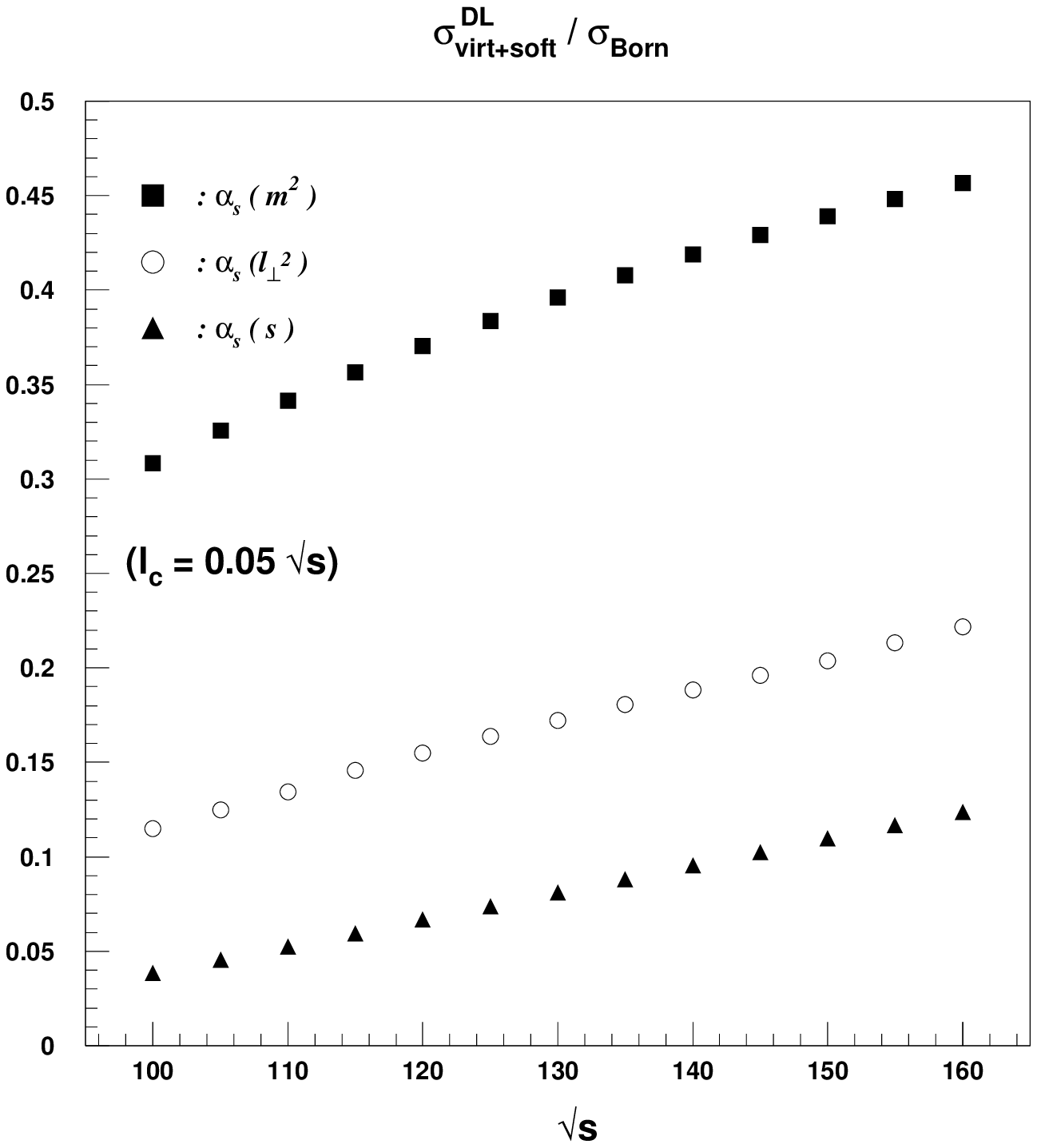,width=10cm}
\caption{The effect of the renormalization group improved form factor (circles)
of Fig. \ref{fig:rg}
in comparison to using the DL form factors with the indicated
values of the strong coupling. The upper plot corresponds to $l_c=0.1 \sqrt{s}$
and the
lower one to $l_c=0.05 \sqrt{s}$. The effect is displayed for the bottom quark
with $m_b=4.5$ GeV.}
\label{fig:vps}
\end{figure}
\end{center}
The effect of the RG-improvement 
lead to
$\sigma^{RG}_{DL}=\sigma_{Born}(1+
{\cal F}^{RG}_{DL}) \exp({\cal F}^{RG}_{Sud})$
and the results are depicted in Fig. \ref{fig:vps} for
two choices of the gluon energy cutoff $l_c\equiv \epsilon \sqrt{s}$ 
compared to the theoretically
allowed upper and lower limits of the DL-approximation evaluated at
$m_b^2$ and $m_H^2$ respectively. In each case the RG-improved result remains
inside the two DL-limits. The effective scale, defined simply as the one
used in the DL-approximation which gives a result close to the RG-improved
values, depends on $\epsilon$, however in general is rather much closer to
$m_q$ than $m_H$ \cite{ms3}.
\begin{center}
\begin{figure}
\centering
\epsfig{file=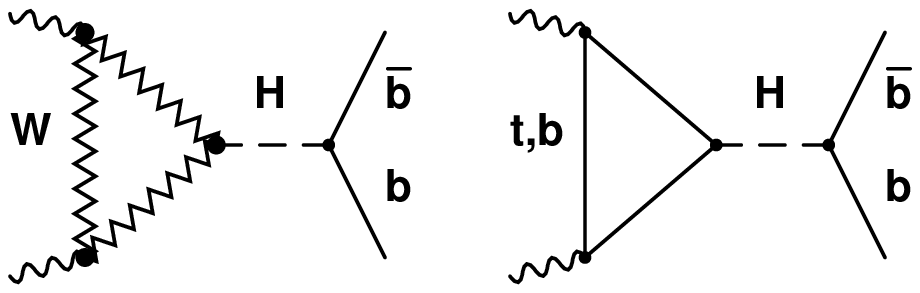,width=12cm} 
\caption{The Standard Model process $\gamma \gamma \longrightarrow H \longrightarrow
b \overline{b}$ is mediated by  $W-$boson and  $t-$ and $b-$quark loops.
}
\label{fig:hgg}
\end{figure}
\end{center}
On the signal side, the relevant radiative corrections have long been known
up to NNL order in the SM \cite{dsbz,dsz} and are summarized including the MSSM
predictions in Ref. \cite{s}. 
For our purposes the one loop corrections
to the diphoton partial width depicted in Fig. \ref{fig:hgg} are sufficient as the QCD corrections
are small in the SM.
The important point to make here and also the novel feature in this analysis is
that the branching ratio BR ($H \longrightarrow b \overline{b}$) is corrected
by the same RG-improved resummed QCD Sudakov form as the continuum heavy quark background
\cite{msk}. This is necessary in order to employ the same two jet definition for
the final state. Since we use the renormalization group improved 
massive Sudakov form factor ${\cal F}^{RG}_{Sud}$ of Ref. \cite{ms3}, we prefer the
Sterman-Weinberg jet definition \cite{sw} schematically depicted in Fig. 
\ref{fig:jet}. We also use an all orders resummed running quark mass evaluated
at the Higgs mass for $\Gamma ( H \longrightarrow b \overline{b})$. For
the total Higgs width, we include the partial Higgs to $b \overline{b}, c
\overline{c}, \tau^+\tau^-, WW^*, ZZ^*$ and $gg$ decay widths with all relevant
radiative corrections.

\section{Numerical Results}

We begin with a few generic remarks concerning the uncertainties in our
predictions. The signal process $\gamma \gamma \longrightarrow H \longrightarrow
b \overline{b}$ is well understood and NNL calculations are available. The
theoretical error is thus negligible \cite{s}.

There are two contributions to
the background process
$\gamma \gamma \longrightarrow q \overline{q}$ which we neglect in this paper.
Firstly, the so-called resolved photon contribution
\cite{dkzz} was found to be a small effect, e.g. \cite{jt}, especially since
we want to reconstruct the Higgs mass from the final two-jet measurements and
impose angular cuts in the forward region. In addition the good charm suppression
also helps to suppress the resolved photon effects as they give the
largest contribution.
The second contribution we do not consider here results from the final state 
configuration where
a soft quark is propagating down the beam pipe and the gluon and remaining
quark form two hard back-to back-jets \cite{bkos}. We neglect this contribution
here due to the expected excellent double b-tagging efficiency and the strong
restrictions on the
allowed acollinearity discussed below\footnote{As discussed in Ref. \cite{bkos}
the B-hadrons from the slow b-quark could be dragged towards the gluon side and
thus give rise to displaced decay vertices in the gluon jet. It may be of inter
est
to perform further systematic MC studies of this effect.}.
A good measure of the remaining theoretical uncertainty in the continuum background
is given by scanning it below and above the Higgs resonance. For precision extractions
of $\Gamma ( H \longrightarrow \gamma \gamma )$ the exact functional form for resonant
energies is still required, though.

In terms of possible systematic errors, the most obvious effect comes from the 
theoretical uncertainty
in the bottom mass determination. Recent QCD-sum rule analyses, however, reach below the 2\% level
for $\overline{m}_b(\overline{m}_b)$ \cite{bs,hg}. For quantitative estimates of exptected systematic experimental
errors it is clearly too early to speculate at this point. The philosophy adopted henceforth is that
we assume that they can be neglected at the 1\% level and concentrate purely on the statistical
error.

We focus here not on specific predictions for cross sections, but instead on the
expected statistical accuracy of the intermediate mass Higgs signal at a PLC.
As detailed in Refs. \cite{bbc}, due to the narrow Higgs width, the signal event
rate is proportional to $N_S \sim \left. \frac{dL_{\gamma \gamma}}{dw} 
\right|_{m_H}$, while the BG is proportional to $L_{\gamma \gamma}$. 

To quantify this, we take the design parameters of the proposed TESLA
linear collider  \cite{t,tp},
which correspond to an integrated peak
$\gamma \gamma$-luminosity of 15 fb$^{-1}$ for the low energy running of the
Compton collider. The polarizations of the incident electron beams and the
laser photons are chosen such that the product of the helicities $\lambda_e
\lambda_{\gamma} = -1$ \footnote{The maximal initial electron polarization
for existing projects is 85 \%, e.g. Ref. \cite{t}.}.
This ensures high monochromaticity and polarization of the photon beams \cite{
t,tp}.
Within this scenario a typical resolution of the Higgs mass is about 10~GeV, so
that for comparison with
the background process $BG \equiv \gamma \gamma \longrightarrow q \overline{q}$
one can use \cite{bbc}
$\frac{L_{\gamma \gamma}}{10\; {\rm GeV}} = \left. 
\frac{d L_{\gamma \gamma}}{dw}\right|_{m_H}$
with $\left. \frac{d L_{\gamma \gamma}}{dw} \right|_{m_H}=$0.5 fb$^{-1}$/GeV.
The number of background events is then given by
$N_{BG} = L_{\gamma \gamma} \sigma_{BG}$.

In principle it is possible to use the exact Compton profile of the backscattered
photons to obtain the full luminosity distributions. The number of expected
events is then given as a convolution of the energy dependent luminosity and
the cross sections. Our approach described above corresponds to an effective
description of these convolutions, since these functions are not precisely
known at present. Note that the functional forms currently used
generally assume that only one scattering takes place for each photon, which may
 not
be realistic.  Once the exact luminosity functions
are experimentally determined  it is of course trivial to
incorporate them into a Monte Carlo program containing the physics described
in this paper.

\begin{center}
\begin{figure}
\centering
\epsfig{file=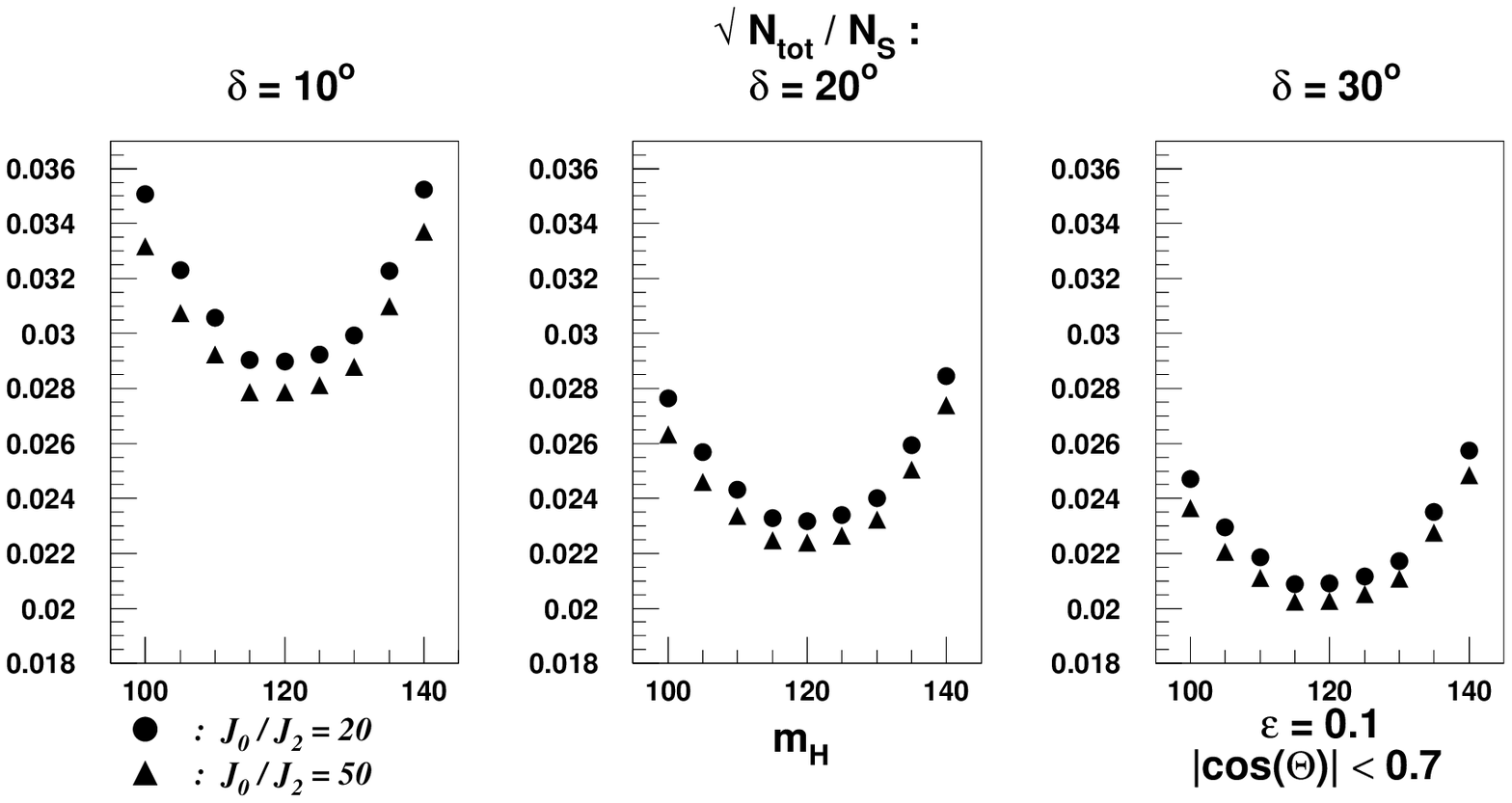,width=17cm}
\epsfig{file=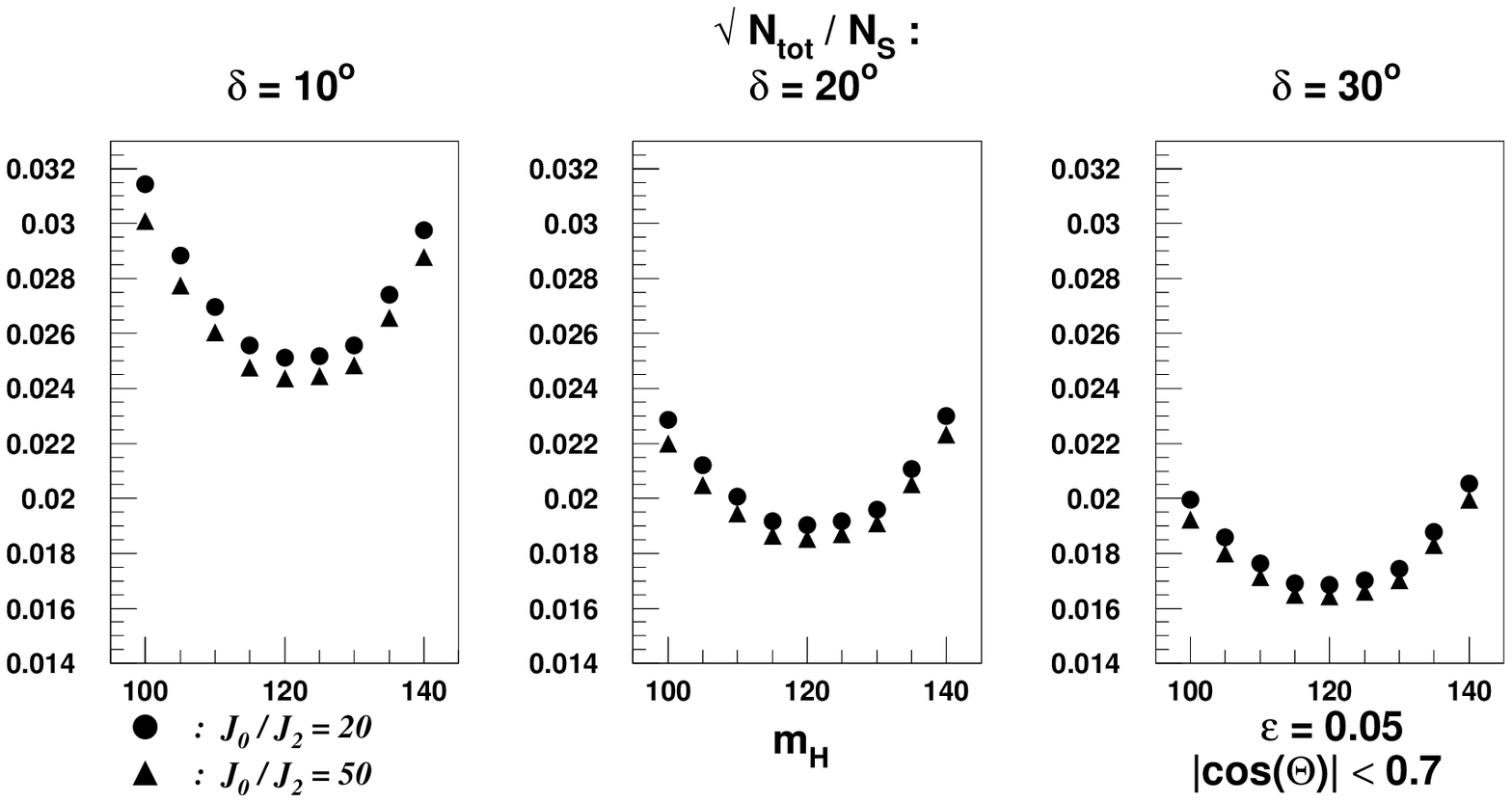,width=17cm}
\caption{The cone-angle dependence of the inverse statistical significance
of the intermediate mass Higgs signal for the displayed values of thrust
and energy cut parameters. Overall a 70\% double b-tagging efficiency and
a 0.5\% charm misidentification rate are assumed. For larger values of $\delta$
the number of events is enlarged, however, the theoretical uncertainty increases
.
For smaller values of $\epsilon$ higher order cutoff dependent terms might becom
e
important.}
\label{fig:ssig}
\end{figure}
\end{center}
In Ref. \cite{msk} it was demonstrated that in order to achieve a large enough
data sample, a central thrust angle cut $| \cos \theta | < 0.7$ is 
advantageous and is adopted here. We also assume a (realistic) 70\%
double b-tagging efficiency. For the charm rejection rate, however,
it seems now possible to assume an even better detector performance.
The improvement comes
from assuming a better single point resolution, thinner detector modules and
moving the vertex detectors closer to the beam-line \cite{ba}.

With these results in hand we keep $| \cos \theta | < 0.7$ fixed
and furthermore assume the $\cc$ misidentification rate of 0.5\%,
(half of that in Ref. \cite{msk}). We vary
the cone angle $\delta$ between narrow ($10^o$), medium ($20^o$) and large
($30^o$) cone sizes for both $\epsilon=0.1$ and $\epsilon=0.05$.
The upper row of Fig. \ref{fig:ssig} demonstrates that for the former
choice of the energy
cutoff parameter we achieve the highest statistical accuracy for the
large $\delta=30^o$ scenario of around 2\%. We emphasize, however, that
in this case also the missing ${\cal O} \left( \alpha_s^2 \right)$ bremsstrahlung
corrections could become important.

The largest effect is obtained by effectively suppressing the background
radiative events with the smaller energy cutoff of $\epsilon=0.05$ outside
the cone (the inside is of course independent of $\epsilon$). Here the
lower row of Fig.
\ref{fig:ssig} demonstrates that the statistical accuracy of the Higgs
boson with $m_H < 130$ GeV can be below the 2\% level after collecting one
year of data. We should mention again that for this choice of $\epsilon$
we might have slightly enhanced the higher order (uncanceled) cutoff dependence.
The dependence on the photon-photon polarization degree is visible but not
crucial. Comparing with the results of Ref. \cite{msk} we also conclude that
the new optimized charm misidentification rate leads to only slight improvements
for $\sqrt{N_{tot}}/N_S$.

In summary, it seems very reasonable to expect that at the Compton collider
option we can achieve a 2\% statistical accuracy of an intermediate mass
Higgs boson signal after collecting data over one year of running.

\section{Conclusions} \label{sec:con}

In this paper we have studied the Higgs signal and continuum background
contributions to the process $\gamma\gamma \longrightarrow
b \overline{b}$ at a high-energy Compton collider.
We have used all relevant QCD
radiative corrections to both the signal and BG production
available in the literature. The Monte Carlo results using a variety of
jet-parameter variations revealed that the intermediate mass
Higgs signal can be expected to be studied with a statistical uncertainty
between (excluding the narrow cone angle scenario) 
$2.4\%$ in a realistic and $1.6\%$ in an optimistic scenario after
one year of collecting data for a Higgs-mass which excludes
$\tan \beta < 2$.

Together with the expected uncertainty of 1\% from the $e^+e^-$ mode determination
of BR$(H \longrightarrow \bb)$, and assuming {\it four} years of collecting data, we
conclude that a measurement of the partial width
$\Gamma (H \longrightarrow \gamma \gamma)$ of 1.4\% precision
level\footnote{This estimate assumes uncorrelated error progression and negligible
systematic errors.}
is feasible for the MSSM mass range from a purely statistical point of view.
This level of accuracy could significantly enhance the kinematical reach
of the MSSM parameter space
in the large pseudoscalar mass limit and thus open up a window for
physics beyond the Standard Model.

For the total Higgs width, the main uncertainty is given by the error in the
branching ratio BR$(H \longrightarrow \gamma \gamma)$, which at present is
estimated at the 15 \% level \cite{brient}. For Higgs masses above 110 GeV,
the total Higgs width could be determined more precisely through the
Higgs-strahlung process \cite{egn,ik} and its decay into $WW^*$ \cite{br}.
This is only possible, however, if the supersymmetric lightest Higgs boson
coupling to vector bosons is universal (i.e. the same for $hWW$ and $hZZ$)
and provided the
optimistic luminosity assumptions can be reached.

In summary, using realistic and optimized machine and detector design parameters, we
conclude that the Compton collider option at a future linear collider
can considerably extend our ability to discriminate between the SM and
MSSM scenarios.

\section*{Acknowledgments}

I would like to thank my collaborators W.J.~Stirling and V.A.~Khoze for their contributions to 
the results presented here. In addition I am grateful for interesting discussions with 
G.~Jikia and V.I.~Telnov.

\end{document}